\documentclass[aps,twocolumn,groupedaddress,showpacs]{revtex4}
\usepackage{longtable}
\usepackage{tabularx}
\usepackage{graphicx}
\usepackage{dcolumn}
\usepackage{bm}
\def\v_op{ \hat{\mathbf v} }

\newcommand{\la}{\langle}
\newcommand{\ra}{\rangle}

\newcommand{\be}{\begin{equation}}
\newcommand{\ee}{\end{equation}}
\newcommand{\bea}{\begin{eqnarray}}
\newcommand{\eea}{\end{eqnarray}}
\newcommand{\etal}{ {\it{et. al.}}}

\renewcommand{\inf}{\infty}

\begin{document}

\title{Auxiliary-field quantum Monte Carlo calculations of 
  molecular systems \\
with a Gaussian basis}

\author{W.~A.~Al-Saidi, Shiwei Zhang, and Henry Krakauer}

\affiliation
{Department of Physics, College of William and Mary,
Williamsburg, VA 23187-8795}

\date{\today}

\begin{abstract}

We extend the recently introduced phaseless auxiliary-field quantum
Monte Carlo (QMC) approach to any single-particle basis, and apply it
to molecular systems with Gaussian basis sets.  QMC methods in general scale favorably with system
size, as a low power.  A QMC approach with auxiliary fields in
principle allows an exact solution of the Schr\"odinger equation in
the chosen basis. However, the well-known sign/phase problem causes
the statistical noise to increase exponentially.  The phaseless method
controls this problem by constraining the paths in the auxiliary-field
path integrals with an approximate phase condition that depends on a
trial wave function.  In the present calculations, the trial wave
function is a single Slater determinant from a Hartree-Fock
calculation. The calculated all-electron total energies show typical 
systematic errors of no more than a few milli-Hartrees compared to
exact results. 
At equilibrium geometries in the molecules we
studied, this accuracy is roughly comparable to that of 
coupled-cluster with single and double excitations and with non-iterative 
triples, CCSD(T).  For stretched bonds in H$_2$O, our
method exhibits better overall accuracy and a more uniform behavior
than CCSD(T).

\end{abstract}

\maketitle

\section{Introduction}

Obtaining the solution of the Schr\"odinger equation, even within
a finite Hilbert space, is an important goal in quantum
chemistry and in condensed matter physics, 
both in its own right and for calibrating 
approximate methods.  In quantum chemistry, 
a variety of approaches have been developed
over the last fifty years, 
which range in quality from the mean-field Hartree-Fock (HF) to the
exact full configuration interaction (FCI) solution. In between these
two limits, a hierarchy of approximations have been developed which
improve upon the Hartree-Fock solution at the expense of increasing
computational cost \cite{qc_ref}.

The FCI method is quite formidable computationally,
because its cost grows exponentially with the number of
electrons and basis functions.
Recently, the density matrix
renormalization group (DMRG) approach was introduced as a
method to potentially obtain near-FCI quality solutions of the Schr\"odinger
equation for quantum chemical Hamiltonians \cite{white_1,dmrg_H2O_41,dmrg_rev}. The results so far are 
promising. For example, DMRG was applied recently \cite{dmrg_H2O_41} to study the water molecule using
a double-zeta atomic-natural-orbital (ANO) basis (41 basis functions), which is currently
intractable with FCI.

Among the approximate approaches, coupled cluster (CC) methods have
been the most successful, especially at equilibrium geometries
\cite{ccsdt_ref,ccsdt_crawford}.  However, standard CC 
methods are non-variational and their computational cost grows rather
steeply with the system size as well.
For example, the most popular among them, CC 
with single and double excitations
and with non-iterative
triples [CCSD(T)],  
has an $N^7$ scaling with the size of the basis. 

Quantum Monte Carlo (QMC) methods are an attractive means for 
a non-perturbative and explicit treatment
of
the interacting many-fermion system. 
These methods tend to have favorable scaling with system size, often 
as a low power.  
The most
established QMC method, the real space fixed-node diffusion Monte Carlo (DMC)
method, 
has been applied successfully to calculate many
properties of solids \cite{QMC_rmp} and molecules \cite{QMC_rev_mole}.

Recently, an alternative and complementary QMC method has been
introduced  \cite{zhang_krakauer}
for realistic electronic Hamiltonians,
based on a field-theoretic approach with auxiliary fields. 
The central idea of auxiliary-field (AF) QMC methods is the
mapping of the interacting many-body problem into a linear combination
of non-interacting problems coupled to external AFs.
Averaging over different AF configurations is performed
by Monte Carlo techniques.
The basic formalism of
AF QMC methods has mostly been applied to lattice models of
strongly interacting systems \cite{BSS,Koonin}.
In these models, the simplified form of the two-body
interactions makes possible 
an efficient mapping with real AFs. As a result, ``only'' a sign 
problem \cite{Zhang_review} occurs.
Constrained path methods \cite{Zhang,Zhang_review} have been 
developed to approximately control the sign problem, which have 
been shown to be quite accurate.

The potential of AF QMC methods for realistic Hamiltonians
has long been recognized and 
pursued \cite{Silvestrelli93,shiftcont_rom97,Wilson95}.
However, 
wide and general applications were not realized because
of a lack of control for the phase problem.
Except for special cases, 
the two-body electronic interactions lead to complex AFs. 
The many-dimensional integration over complex
variables in turn leads to an
exponentially increasing noise and therefore breakdown of the method.
The phaseless AF QMC method \cite{zhang_krakauer} was introduced to
control the phase problem in an approximate manner. 

In Ref.~\cite{zhang_krakauer}, a general framework was developed on
how to use importance
sampling of the determinants to deal with the complex phase. An
approximate constraint was then formulated with a trial wave function to
constrain the paths in AF space and eliminate the growth
of noise. 
Because of the constraint, the calculated ground state energy is no longer
exact and is not
variational.  In applications to
several $sp$-bonded materials \cite{zhang_krakauer,zhang_krakauer2,cherry},
it was shown that the method, with a plane-wave basis and simple trial wave
functions, gave accurate results for
systems from simple atoms to large supercells. 
The phaseless AF QMC method was also recently
applied to the strongly correlated transition metal oxide
molecules TiO and MnO \cite{alsaidi}, again yielding 
results in good agreement with experiment using simple mean-field trial
wave functions.

The success of the phaseless AF approach in
solid-state applications with plane-waves has motivated us to extend it
to a generic one-particle basis, targeting in particular
quantum chemistry problems.  
Applying the new AF QMC method to such problems is very
appealing, especially with 
the integral role basis sets play in 
quantum chemistry methods. 
This method allows
QMC calculations using any choice of one-particle basis.
It provides a framework for general many-body 
calculations which can import straightforwardly 
many of the well-established technologies from more traditional approaches.
By importance-sampled
random walks,
the phaseless AF QMC method obtains the many-body ground state 
with an ensemble of loosely coupled independent-particle calculations
which propagate with imaginary-time.
The ground-state wave function is represented by a 
linear combination of non-orthogonal Slater determinants.

In this paper we report the first results from this effort. We discuss
aspects of the new method when implemented with Gaussian basis sets.
We present benchmark results on $sp$-bonded atoms and molecules.
All-electron total energies are calculated for various first-row atoms
and molecules, and compared with FCI and DMRG results where available
or with CCSD(T)
otherwise. The behavior of the method
is also studied as a function of basis size, including an ANO basis
\cite{roos,roos_footnote1} in
H$_2$O and O$_2$, each with $92$ basis functions.
We then test the method away
from equilibrium geometries, calculating the equilibrium bond length
and also stretching the bonds in the water molecule by a factor
between $1$ and $8$ within a cc-pVDZ basis \cite{cc_basis}. 
In all our calculations the trial
wave function is taken to be a single Slater determinant from
Hartree-Fock, either restricted (RHF) or unrestricted (UHF). Their
effect on the accuracy is examined.  
The calculated energies with the optimal HF wave function typically show systematic
errors of no more than a few milli-Hartree (mE$_{\rm h}$) compared to
exact results.  For stretched bonds in H$_2$O, our method exhibits
better overall accuracy and more uniform behavior than CCSD(T).

An additional benefit of the present AF QMC implementation 
is its value for algorithmic studies. 
Compared to the plane-wave algorithm, the generic basis AF QMC method 
typically has much more effective basis sets. As a result, the basis
size is often much smaller and the method is significantly
cheaper computationally. Further,
direct comparisons can be more easily
made between the QMC results and those obtained with more established
correlated methods, as indicated above. 
Extra ingredients
such as pseudopotentials can be removed, with the only remaining
uncertainty being the systematic error from the phaseless approximation. 
In this study, we take advantage of this feature to make detailed
benchmark studies of the systematic error. 

The rest of the paper is organized as follows. The phaseless AF QMC
method is first briefly reviewed in the next section. We then
describe its implementation with the Gaussian basis in 
Sec.~\ref{sec:implement}, together with some results to illustrate
the behavior and characteristics of this approach. Section~\ref{sec:computat_details}
gives some of the details of the computational procedures.
In Sec.~\ref{sec:equil-geometries} and Sec.~\ref{sec:H2O-bondstretch}, we
present the results of our calculations, including calculations of 
the H$_2$O  molecule both 
at the equilibrium geometry and with stretched bond lengths,
and 
total energies
and energy differences (binding, ionization, and electron 
affinity) for various other first-row atoms and diatomic molecules.
Finally
in Sec.~\ref{sec:summary} we make concluding remarks, together
with a brief discussion of possibilities for further improvement of
the new algorithm.

\section{Formalism }
\label{sec:formalism} 

The full electronic Hamiltonian for a many-fermion system with two-body interactions
can be written in any orthonormal one-particle basis in the general form
\begin{equation}
{\hat H} ={\hat H_1} + {\hat H_2}
= \sum_{i,j}^N {T_{ij} c_i^\dagger c_j}
   + {1 \over 2} 
\sum_{i,j,k,l}^N {V_{ijkl} c_i^\dagger c_j^\dagger c_k c_l},
\label{eq:H}
\end{equation}
where $N$ is the size of the chosen one-particle basis, and
$c_i^\dagger$ and $c_i$ are the corresponding creation and
annihilation operators.  Both the one-body ($T_{ij}$) and two-body ($V_{ijkl}$) 
matrix elements  are known.

The auxiliary-field quantum
Monte Carlo method obtains  the ground state $\left| \Psi_G \right\rangle$ of
${\hat H}$, by projecting from a trial wave function $\left| \Psi_T
\right\rangle$ using the imaginary-time propagator $e^{-\tau {\hat
H}}$:
\be
\left| \Psi_G \right\rangle \propto  \lim_{n
\to \infty} (e^{-\tau {\hat H}})^n \left| \Psi_T \right\rangle \label{eq:proj}. 
\ee
The trial wave function $\left| \Psi_T \right\rangle$, which should
have  a non
zero overlap with the exact ground state, is assumed to be
in the form of a single determinant or a linear combination of
Slater determinants.

The timestep, $\tau$, is chosen to be small enough so that ${\hat H_1}$ and
${\hat H_2}$ in the propagator can be accurately separated with the
Trotter decomposition:
\begin{equation}
e^{-\tau {\hat H}} = e^{-\tau {\hat H_1}/2} e^{-\tau {\hat H_2}}
e^{-\tau {\hat H_1}/2} + {\cal{O}}(\tau^3).\label{eq:expH}
\end{equation}
The action of the propagator $e^{-\tau {\hat H_1}}$ on a Slater
determinant yields another determinant.  This  is not the case with
$e^{-\tau {\hat H_2}}$, which however can be written as an integral of 
one-body operators using a Hubbard-Stratonovich (HS) transformation \cite{HS}: 
\begin{equation}
   e^{-\tau{\hat H_2}}
= \prod_\alpha \Bigg({1\over \sqrt{2\pi}}\int_{-\infty}^\infty
d\sigma_\alpha \,
            e^{-\frac{1}{2} \sigma_\alpha^2}
           e^{\sqrt{\tau}\,\sigma_\alpha\,
\sqrt{\zeta_\alpha}\,{\hat v_\alpha}} \Bigg),
\label{eq:HStrans1}
\end{equation}
where the {\emph{one-body operators}} ${\hat v_\alpha}$ can be defined generally
for any two-body operator by writing the latter in a quadratic form, such as  ${\hat H_2} = - {1\over 2}\sum_\alpha
\zeta_\alpha {\hat v_\alpha}^2$, with  $\zeta_\alpha$  a real
number.
Monte Carlo methods are very efficient at evaluating  multi-dimensional
integrals as in Eq.~(\ref{eq:HStrans1}). For example, the projection of
Eq.~(\ref{eq:proj}) can be realized iteratively: An ensemble of Slater
determinants $\{\,|\phi \ra\,\}$ are initialized to the trial wave
function $|\Psi_T\ra$, which are then propagated to a new ensemble
$\{\,|\phi'\ra\,\}$ using  $e^{-\tau {\hat H}}$ of Eq.~(\ref{eq:expH}), 
and so on, until convergence is reached. For each Slater determinant 
in the ensemble, the two-body part in Eq.~(\ref{eq:HStrans1}) 
can be applied stochastically by sampling an AF configuration,
$\{\,\sigma_\alpha\,\}$. In the standard AF QMC approach, 
the projection is often 
done \cite{Zhang_review}
as a path integral with paths of fixed length in imaginary time, 
using a Metropolis or heat-bath algorithm.

This straightforward approach, however, will generally lead to an
exponential increase in the statistical fluctuations with the
projection time.  The source of the fluctuations is that
the one-body operators $\v_op \equiv \{\,{\hat v_\alpha}\,\}$ are generally complex,
since usually $\zeta_\alpha$ cannot all be made
positive in Eq.~(\ref{eq:HStrans1}).  As a result, the orbitals in
$|\phi\rangle$ will become complex as the projection proceeds. 
For large projection time, the phase of
each $|\phi\rangle$ becomes random, and the MC representation of
$|\Psi_G\rangle$ becomes dominated by noise. This
leads to the phase problem and the divergence of the fluctuations. The
phase problem is of the same origin as the sign problem that occurs
when the one-body operators $\v_op$ are real, but 
is more severe because for each
$|\phi\rangle$, instead of a $+|\phi\rangle$ and $-|\phi\rangle$
symmetry \cite{Zhang}, there is now an infinite set 
$\{ e^{i\theta} |\phi\rangle, \theta \in [0,2\pi) \}$ among which the Monte Carlo
sampling cannot distinguish.

We used the phaseless auxiliary-field QMC method to control the phase
problem. In order to implement a phaseless constraint, 
the method recasts the imaginary-time path integral as a branching random walk
in Slater-determinant space \cite{Zhang,zhang_krakauer}.
It uses a trial wave
function $|\Psi_T\ra$ and a {\em complex} importance 
function, $\langle \Psi_T|\phi\rangle$, to construct phaseless
random walkers, $|\phi\ra/\langle \Psi_T|\phi\rangle$, which are invariant
under a phase gauge transformation. The resulting two-dimensional diffusion
process in the complex plane of the overlap $\langle
\Psi_T|\phi\rangle$ is then approximated as a diffusion process in one
dimension. 

As mentioned earlier, the ground-state energy computed with the so-called mixed estimate 
is approximate and not variational in the phaseless method.
The error depends on $|\Psi_T\rangle$,
vanishing when $|\Psi_T\rangle$ is exact. This is the only error in the method that cannot 
be eliminated systematically. 
The method has been applied to atoms, 
molecules, and simple solids, using a plane-wave basis and pseudopotentials,
and has proved very successful 
\cite{zhang_krakauer,zhang_krakauer2,alsaidi}.

\section{Implementation using a localized basis}
\label{sec:implement}

With Gaussian basis sets, the matrix elements of interest to QMC and
the overlap matrix can be imported from quantum chemistry programs.
It is more convenient to use orthonormal basis
functions, which ensure the usual commutation relations of the
creation and destruction operators.  To set up the Hamiltonian in
Eq.~(\ref{eq:H}), we first transform the non-orthogonal basis into an
orthogonalized basis set.
The one- and two-body matrix elements,
$T_{ij}$ and $V_{ijkl}$, are then expressed with respect to this basis 
via the transformation, based on the original matrix elements. 

To carry out the HS transformation of 
Eq.~(\ref{eq:HStrans1}), we first map the two-body interaction matrix
$V_{ijkl}$ into a real, symmetric supermatrix ${\cal{V}}_{\mu[i,l],\nu[j,k]}$
where $\mu, \nu= 1,\ldots,N^2$. This is then expressed in terms of its
eigenvalues $(-\lambda_{\alpha})$ and eigenvectors $X_{\mu}^{\alpha}$: $
{\cal{V}_{\mu,\nu}}= -\sum_{\alpha} \lambda_{\alpha}
X_{\mu}^{\alpha}\,X_{\nu}^{\alpha} $.

The two-body operator ${\hat H}_2$ of Eq.~(\ref{eq:H}) can be written as the
sum of a one-body operator ${\hat H}'_{1}$ and a two-body
operator ${\hat H}'_{2}$. The latter can be further expressed in terms of the
eigenvectors of ${\cal{V}}_{\mu,\nu}$ as 
\be 
{\hat H'}_2= -\frac{1}{2}\,\sum_{\alpha} \lambda_{\alpha} {\hat v^2}_{\alpha},  \label{eq:hs_decomp} 
\ee
where the one-body operator ${\hat v}_{\alpha}$ is, 
\be
{\hat v}_{\alpha}= \sum_{i,l} X_{\mu[i,l]}^{\alpha} a^{\dag}_{i} a_{l}.
\ee
Note that ${\hat v^\dagger}_{\alpha}= {\hat v}_{\alpha}$ for real basis
functions, as is the case here. The form in Eq. (5) is now  amenable to the  HS transformation of
Eq.~(\ref{eq:HStrans1}). 

In this case the number of the HS fields is equal to the number
of non-zero eigenvalues $(- \lambda_\alpha)$ of the symmetrized
two-body supermatrix.  Other forms of decomposition and HS
transformation are possible, and their efficiency and effectiveness
with a constraint can vary.(For example, using a modified Cholesky
decomposition will eliminate the need to diagonalize the supermatrix
and potentially reduces the computational cost of synthesizing the matrix. However,
this will generally lead to a larger number of HS fields, and hence increase the
QMC computational cost.) We will defer further analysis
of this freedom to a future publication.

\begin{figure}[t]
\includegraphics[width=8 cm]{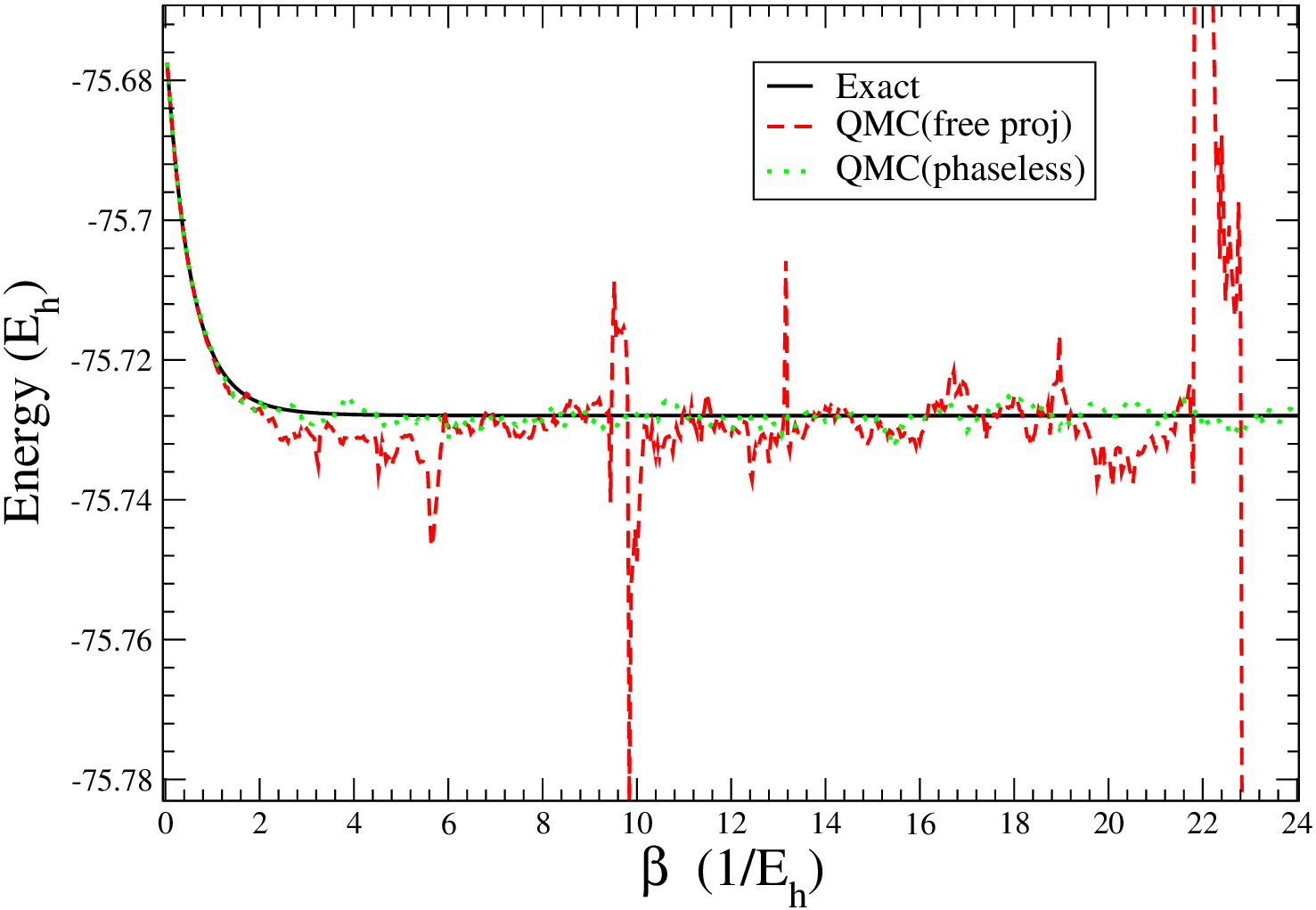} 
\caption{ The projection of the ground state of H$_2$O/STO-6G. Starting from the
  HF trial wave function, the total energy is plotted vs.~the projection time
  $\beta=n\,\tau$, as
  obtained using the free projection QMC (exact but suffers from the
  phase problem) and the phaseless approximation. For
  comparison, we show also the exact projection which is obtained form
  diagonalization of the many-body Hamiltonian.  
  The phaseless QMC method tracks smoothly the exact
  projection with no sign of the uncontrolled fluctuations in free projection.
}
\label{fig_proj}
\end{figure}

To illustrate the method, we use the
H$_2$O molecule in a minimal STO-6G basis. This problem is small enough
to permit detailed comparison with exact diagonalization. 
Starting from HF solution, 
the projection in Eq.~(\ref{eq:proj}) was applied using: i) the exact many-body
propagator $e^{-\tau \hat H}$ acting on the wave function expanded in terms of
the exact eigenstates of the many-body Hamiltonian ${\hat H}$, ii) the free projection QMC
method (no constraint imposed) which is exact but suffers from the phase problem, 
and iii) the
phaseless QMC which is approximate but controls the phase
problem. Figure~\ref{fig_proj} plots the results of total-energy mixed-estimator \cite{zhang_krakauer},
corresponding to these three calculations. 
Up to a projection time of $\beta~\approx~2$~E$_h^{-1}$,
both QMC methods show essentially the same convergence of the total-energy. For large
projection times, the free-projection starts showing the
phase problem in the form of large fluctuations. Around
$\beta~\approx~23$~E$_h^{-1}$  these fluctuations diverge on the scale of the plot. For larger
number of particles or larger basis sets, this divergence would have occurred at much
earlier projection times $\beta$. The phaseless QMC method, by contrast, follows the
exact projection with finite variance as the random walk proceeds.

In our implementation, we found it advantageous to first subtract the
mean-field ``background'' from ${\hat v}_{\alpha}$ before applying the
Hubbard-Stratonovich transformation. In this case,
Eq.~(\ref{eq:hs_decomp}) is replaced by,
\be
 {\hat H}'_2= -\frac{1}{2} \sum_{\alpha} \lambda_{\alpha} 
({\hat v}_{\alpha}-\la{\hat v}_{\alpha}\ra)^2 
+ {\hat H''}_1, 
\label{eq:hs_decomp_shift}
\ee
where $\la{\hat v}_{\alpha}\ra=\la \Psi_T|{\hat v}_{\alpha}|\Psi_T\ra$ and ${\hat H''}_1$ is the one-body operator:
\be
{\hat H''}_1= -\frac{1}{2} \, \sum_{\alpha}\left( 
{\hat v}_{\alpha} \la{\hat v}_{\alpha}\ra
+ {\hat v}_{\alpha} \la{\hat v}_{\alpha}\ra
- \la{\hat v}_{\alpha}\ra \la{\hat v}_{\alpha}\ra \right).
\ee
Equations~(\ref{eq:hs_decomp}) and
(\ref{eq:hs_decomp_shift}) are equivalent.
When the approximate phaseless projection is imposed, however, 
using Eq.~(\ref{eq:hs_decomp_shift}) before
applying the HS transformation leads to an improved behavior.

\begin{figure}[t]
\includegraphics[width=8cm]{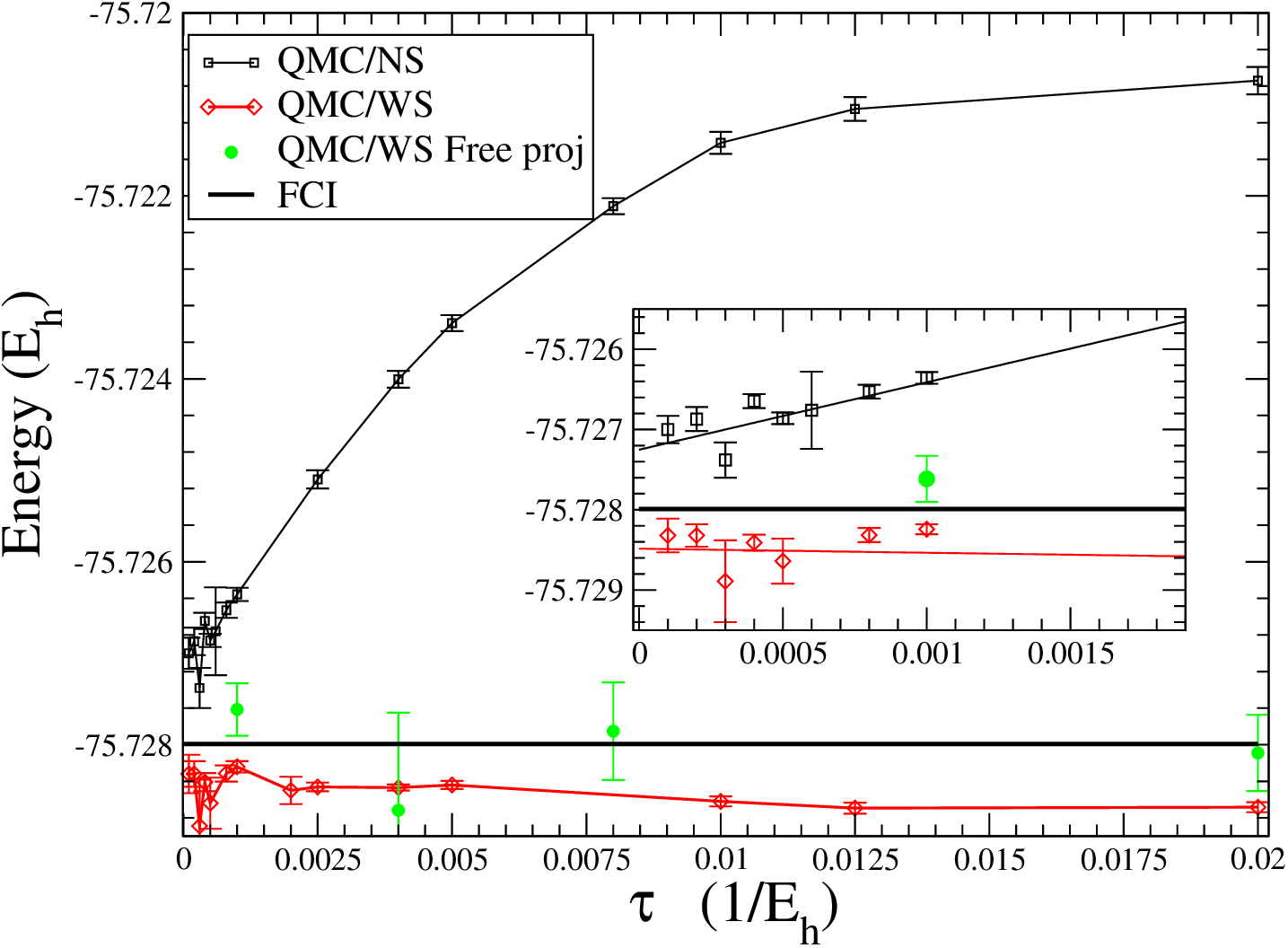} 
\caption{ QMC Trotter time-step dependence. The total energy of the
H$_2$O/STO-6G  molecule from phaseless QMC is shown vs.~the time-step size
$\tau$, using the HS transformations of
Eq.~(\ref{eq:hs_decomp}) (labeled as QMC/NS) and of
Eq.~(\ref{eq:hs_decomp_shift}) (labeled as QMC/WS, in which the
mean-field ``background'' is subtracted). Results from
free-projection (exact but suffers from the phase problem) are also
shown, together with that of the exact FCI. Points are connected by
straight lines for clarity.  The insert shows a closer look at small
values of $\tau$, together with linear fits of the QMC data.
}
\label{fig_shift}
\end{figure}

We illustrate this point in Fig.~\ref{fig_shift}. 
Using the same STO-6G minimal basis, we 
plot the Trotter time-step convergence of the total energy of the H$_2$O
molecule, calculated with the HS transformations of
Eq.~(\ref{eq:hs_decomp}) and Eq.~(\ref{eq:hs_decomp_shift}), for both
the phaseless QMC method and for (exact) free-projection.  For comparison,
the ground state energy obtained from exact diagonalization is also
shown.  
The Trotter extrapolated QMC energy (see inset of
Fig.~\ref{fig_shift}) is $-75.72725(4)$~E$_h$ with the HS
decomposition of Eq.~(\ref{eq:hs_decomp}) and $-75.72849(5)$~E$_h$
with Eq.~(\ref{eq:hs_decomp_shift}). Monte Carlo statistical error
bars are on the last digit and are shown in parentheses. For comparison,
the exact ground-state energy is $-75.72799$~E$_h$.

Similar results and analysis have also been presented in phaseless AF QMC calculations in boson
systems \cite{Wirawan05}. 
We comment that the issue in Fig.~\ref{fig_shift} is different from 
the shifted contour approach \cite{shiftcont_rom97},
dispite formal similarities.
With the importance sampling scheme of Ref.~\cite{zhang_krakauer}, 
Eqs.~(\ref{eq:hs_decomp}) and
(\ref{eq:hs_decomp_shift}) would both be exact and give the same results
under free projection, since the mean-field shift is 
automatically applied via the importance sampling 
transformation regardless of which form of ${\hat H}'_2$ is used 
\cite{Wirawan04}. 
The different behaviors discussed above arise only because 
of the imposition of the phase projection to 
one-dimension \cite{zhang_krakauer}, in which the substraction of the 
mean-field background helps to reduce 
the ``rotation'' of the random 
walkers in the complex $\langle\Psi_T|\phi\rangle$-plane and thus 
the severity of the projection.

\section{Computational Details}
\label{sec:computat_details}
We apply our method to several atomic and molecular systems,
which were chosen primarily for benchmarking purposes and which have all
been previously studied using well-established correlated methods such
as coupled cluster (CC), full configuration interaction (FCI), or
density matrix renormalization group (DMRG) methods. In these and in our QMC
calculations,  the core and
valence electrons are fully correlated. 

All the QMC calculations reported below used a single Hartree-Fock
Slater determinant as the trial wave function. 
For all systems, QMC was done with  the
best variational single determinant, namely the unrestricted
Hartree-Fock (UHF) solution. In addition, we have also tested the
restricted Hartree-Fock (RHF) solution as the trial wave function in some
cases. 
No additional optimization or tuning of the trial wave
function was performed beyond the mean-field Hartree-Fock calculation.

The QMC code is interfaced with GAMESS\cite{gamess} and
NWChem\cite{nwchem} to import the Gaussian one-electron and
two-electron matrix elements, the overlap matrix, and the trial wave
function. All of our calculations are done using the
spherical harmonics representation of basis functions.

We performed the FCI calculations
using MOLPRO~\cite{molpro,fci_molpro} and the  coupled cluster calcuations using
NWChem~\cite{nwchem}.  Unless otherwise noted, the coupled cluster
calculations for atoms and molecules with a singlet state are of the
type RHF-RCCSD(T) i.e. based on the RHF reference state,
while those for a multiplet state are of the unrestricted type,
UHF-UCCSD(T) based on the UHF solution. 

\begin{table}[t]
\caption{The binding energy (BE) of H$_2$O calculated with four basis
sets: STO-6G, cc-pVDZ, and double- and triple-zeta ANO
\cite{roos_footnote1}.  Also shown are all-electron total energies for
O, H, and H$_2$O.  Monte Carlo statistical errors are in the last
digit and are shown in parentheses.  All energies are in Hartrees. The FCI
value for H$_2$O/cc-pVDZ is from Ref.~\cite{H2O_Olsen}, while the
DMRG energy of H2O and the FCI energy of O within the double-zeta ANO
are from Ref.~\cite{dmrg_H2O_41}.}
\begin{tabular}{l c c c l}
\hline
\hline
      &  UHF &  CCSD(T) & FCI/DMRG  &   \ \ QMC \\
\hline
STO-6G\\
H     & \ -0.471\,039 &\  -0.471\,039 &\  -0.471\,039 &\  -0.471\,039 \\
O     &  -74.516\,816 &  -74.516\,816 &  -74.516\,816 &  -74.516\,816  \\
H$_2$O   &  -75.676\,506 &  -75.727\,931 &  -75.727\,991 &  -75.728\,5(1)     \\
BE    &\ \ 0.217\,612 &\ \ 0.269\,037 &\ \ 0.269\,097 &\ \ 0.269\,5(1) \\
\\
 cc-pVDZ \\ 
H     & \ -0.499\,278 &\  -0.499\,278 &\  -0.499\,278  &\  -0.499\,278  \\
O     &  -74.792\,166 &  -74.911\,552 &  -74.911\,744                                        
                                              & -74.909\,6(1) \\
H$_2$O   &  -76.024\,039 &  -76.241\,201 &  -76.241\,860 
                                              & -76.242\,4(2)         \\
BE    &\ \ 0.233\,317 &\ \ 0.331\,093 &\ \ 0.331\,560  &\ \ 0.334\,3(2) \\
\\
DZ ANO\\
H   & \ -0.499\,944 &\  -0.499\,944 &\  -0.499\,944  &\  -0.499\,944  \\
O   &  -74.816\,273 & -74.961\,956  & -74.962\,350
                                         & -74.959\,6(1) \\
H$_2$O &  -76.057\,621 & -76.314\,141  & -76.314\,715 
                                             & -76.316\,3(6)      \\
BE  &\ \ 0.241\,460 &\ \ 0.352\,959 &\ \ 0.352\,477  &\ \ 0.356\,8(7)\\
\\
TZ ANO\\
H   &\  -0.499\,973 &\  -0.499\,973   &           &\  -0.499\,973     \\
O   &  -74.818\,648 &  -75.000\,129    &           & -74.997\,0(4)  \\
H$_2$O &  -76.060\,589 &  -76.367\,528   &           & -76.370(1)      \\
BE  &\ \ 0.241\,995 &\ \ 0.367\,453   &           &\ \ 0.373(1)        \\
\hline
\hline
\end{tabular}
\label{table_H2O_bind}
\end{table}

\section{Equilibrium Geometries}
\label{sec:equil-geometries}
As a first test of our method, we calculate in this section total energies and binding
energies for some well-studied systems at their equilibrium
geometries.  The first subsection contains results for the water
molecule in the vicinity of the equilibrium structure and the O$_2$ dimer 
at equilibrium bond length.  The second subsection contains results for other first-row
atoms and diatomic molecules, including total ground-state energies,
binding energy, convergence with basis sets, atomic ionization
potentials, and electron affinities.  In the next section, we test the method by studying 
bond stretching of the water
molecule.

\begin{figure}[t]
\includegraphics[width=8.5cm]{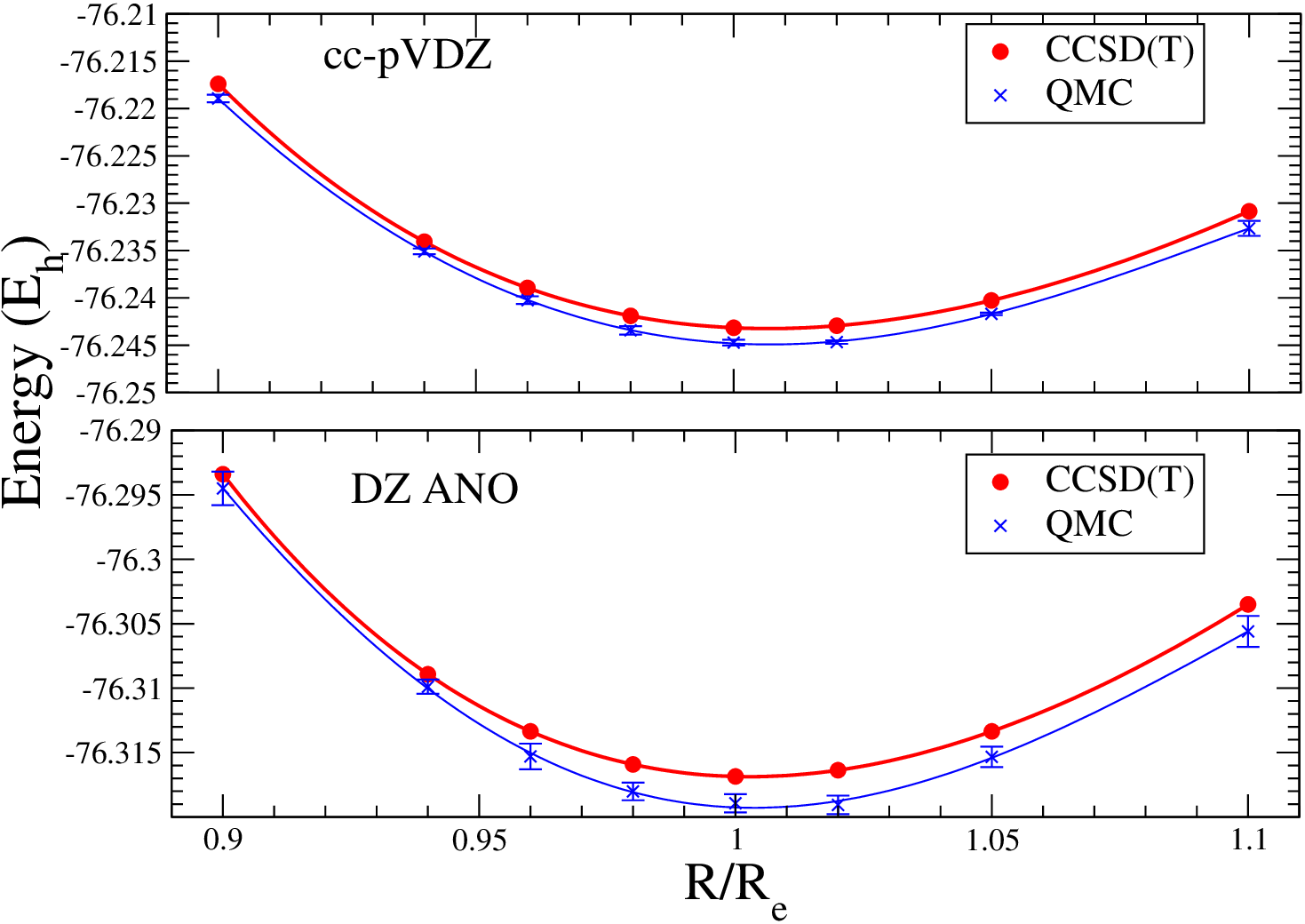}
\caption{
Calculation of the equilibrium bond length of H$_2$O within the
cc-pVDZ basis and DZ ANO basis, using
CCSD(T) and the present QMC method. 
All-electron total energies (in Hartrees) are shown as a function of the O-H bond length
(in units of the experimental equilibrium value, $R_e$), with the angle
between the two O-H bonds fixed at the experimental value.
Monte Carlo statistical error bars are indicated. The lines are based
on a polynomial fit to the data.}
\label{fig:H2O_geom}
\end{figure}

\subsection{H$_2$O and O$_2$}

The water molecule has a long benchmarking history
\cite{saxe,Bauschlicher,H2O_Olsen,H2O_DMC,dmrg_H2O_41}.
Table~\ref{table_H2O_bind} presents results calculated by HF, CCSD(T),
FCI, and the present QMC method.  We show the all-electron total
energies of 
the individual atoms H and O, and the total and binding energies of H$_2$O.
Four different basis sets were used: minimal STO-6G, cc-pVDZ
\cite{cc_basis}, and the double zeta (DZ) and triple-zeta (TZ) ANO
bases of Widmark, Malmqvist, and Roos \cite{roos,roos_footnote1}. For H$_2$O,
these have 7, 24, 41, and 92 basis functions, respectively.  
The H$_2$O
geometries were as follows: results using the minimal \mbox{STO-6G}
correspond to positions (in Bohr) of O$(0, 0, 0)$ and H$(0, \pm
1.515263, -1.058898)$ with $C_{2v}$ symmetry, while the cc-pVDZ
results used the geometry of Ref.~\cite{H2O_Olsen}. The DZ ANO and TZ ANO
results were obtained with the geometry given in
Ref.~\cite{dmrg_H2O_41}.  

The QMC calculations used UHF trial wave
functions. (For H$_2$O, the RHF and UHF solutions are
identical at the equilibrium bond length.) 
The FCI total energies of H$_2$O/cc-pVDZ are those of
Ref.~\cite{H2O_Olsen}.  FCI results for H$_2$O are not within reach
for the DZ ANO basis, but DMRG results are available from a recent
study \cite{dmrg_H2O_41}, which are shown under FCI.  The FCI energy
of O/DZ ANO is also from Ref.~\cite{dmrg_H2O_41}. Neither FCI nor DMRG is
within reach presently for the TZ ANO basis.

\begin{table}[t]
\caption{ Binding energy (BE) of O$_2$ using cc-pVDZ, double- and
  triple-zeta ANO \cite{roos_footnote1} basis sets.  The calculated
  all-electron total energies are also shown.  The experimental bond
  length of $R_e=1.2074$\,\AA \, was used.  O atom results are in
  Table~\ref{table_H2O_bind}.  Monte Carlo statistical errors are in
  the last digit and are shown in parentheses.  All energies are in
  Hartrees. } 
\begin{tabular}{l c c c } 
\hline \hline & UHF & UCCSD(T)  & QMC \\ 
\hline 
cc-pVDZ\\
O$_2$&   -149.627\,780 &   -149.989\,319  &   -149.982\,6(3)\\
BE   &\ \ \ 0.043\,448 &\ \ \ 0.166\,215  &\ \ \ 0.163\,5(3) \\
\\
DZ ANO\\
O$_2$&   -149.679\,669 &   -150.095\,668  &   -150.090\,7(6) \\
BE &  \ \ \ 0.047\,123 &\ \ \ 0.171\,756  &\ \ \ 0.171\,5(7) \\
\\
TZ ANO\\
O$_2$&   -149.689\,251 &   -150.187\,965 &      -150.186(2)   \\
BE   &\ \ \ 0.051\,955 &\ \ \ 0.187\,706 & \ \ \   0.192(2)   \\
\hline
\hline
\end{tabular}
\label{table_O2_bind}
\end{table}

\begin{table*}
\caption{The calculated all-electron total energies of various
  first-row atoms and diatomic molecules in their ground states. We
  show the RHF, UHF, CCSD(T) energies, and the FCI results where
  available.  The basis set and the number of basis
  functions are shown in the third column. Our QMC results based on
  the RHF and UHF trial wave functions are shown in two
  columns. MC statistical errors are in the last digit and shown in
  parentheses. For completeness, we show in the last column some of
  the available estimated experimental, non-relativistic, infinite nuclear mass
  energy (i.e., the estimated exact result at the infinite basis size
  limit).  All energies are in Hartrees.}
\begin{tabular}{l c c c c c c c c l}
\hline
\hline
 & R$_e$(\AA) & basis(\#) &  RHF &  UHF &  CCSD(T)&  FCI & QMC/RHF &  QMC/UHF & E$_{\rm{est}}^{\inf}$ \\
\hline
H$_2$ & 0.741 & cc-pVDZ(10) &$-$1.128\,714 &   &  $-$1.163\,411  &
$-$1.163\,411 & $-$1.163\,60(1) & & $-$1.174\,4\\
Li       &     & cc-pVDZ(14) &  &  $-$7.432\,421 & $-$7.432\,638  &
$-$7.432\,638&  & $-$7.432\,633(3) & $-$7.478\,1\\
Li$_2$ &2.673& cc-pVDZ(28) & $-$14.869\,499 &  $-$14.870\,128  &
$-$14.901\,386& $-$14.901\,392 & $-$14.902\,5(1)& $-$14.900\,35(9) & $-$14.995\,4 \\
Be & &  cc-pVDZ(14) & $-$14.572\,338   &   $-$14.572\,611    &
$-$14.617\,407&$-$14.617\,409  & $-$14.617\,26(7) & $-$14.617\,63(8) & $-$14.667\,4 \\
Be$_2$ &2.45 & cc-pVDZ(28) &  $-$29.132\,211  & $-$29.154\,535  &
$-$29.234\,246&   & $-$29.231\,3(2) & $-$29.234\,1(2) & $-$29.338\,5 \\
Be & &  cc-pVTZ(30) & $-$14.572\,874 &  $-$14.573\,183  & $-$14.623\,790&
$-$14.623\,810   & $-$14.622\,4(2) & $-$14.622\,8(2)& $-$14.667\,4 \\
Be$_2$ & 2.45 & cc-pVTZ(60) &$-$29.133\,688 & $-$29.156\,165  &
$-$29.253\,734& & $-$29.248\,8(5) &  $-$29.251\,1(3) &  $-$29.338\,5 \\
N        &     & cc-pVDZ(14) &  & $-$54.391\,115  & $-$54.479\,944  &  $-$54.480\,115 & & $-$54.479\,56(7)&$-$54.589\,3 \\
N$_2$  & 1.094& cc-pVDZ(28)&  $-$108.954\,553 &   &$-$109.278\,722&
&$-$109.281\,6(3)& & $-$109.542\,3 \\  
F        &    & cc-pVDZ(14) &  & $-$99.375\,240 &  $-$99.529\,322  &   $-$99.529\,518&  & $-$99.528\,1(3)&$-$99.734  \\
F$_2$    &1.412 & cc-pVDZ(28) & $-$198.685\,670 & $-$198.695\,746
&$-$199.101\,151& & $-$199.100\,3(5)  & $-$199.102\,4(4)  &  \\
\\
BH & 1.256&  cc-pVDZ(19) &  $-$25.125\,188 &$-$25.131\,227 & $-$25.215\,917
&   $-$25.216\,401 & $-$25.217\,5(3) & $-$25.215\,1(2) & \\
CH$^+$ & 1.146& cc-pVDZ(19) &  $-$37.900\,480 &
$-$37.909\,823 &  $-$38.003\,207 &
$-$38.003\,712\footnote{Ref. \cite{abrams}} & $-$38.006\,9(3) &
$-$38.000\,5(4) & \\
NH & 1.056 &  cc-pVDZ(19) &   &
$-$54.966\,091 & $-$55.093\,298&
$-$55.093\,721\footnote{Ref. \cite{abrams}} &     &  $-$55.093\,18(7) &  \\
OH$^+$ &1.032&  6$-$31G$^{**}$(19) &  &  $-$74.973\,345 &  $-$75.093\,371&
&     & $-$75.093\,08(9) &  \\
HF &  0.920  & cc-pVDZ(19) &  $-$100.019\,280 &  &  $-$100.230\,098&   &  $-$100.231\,2(1) &   &  \\
\hline
\hline
\end{tabular}
\label{summary_molecules}
\end{table*}

We see that the agreement between the QMC, CCSD(T), and FCI results is
quite good. As mentioned, the calculated ground-state energies in the
present phaseless QMC method are not variational, which can be seen in
the results in H$_2$O, for example.  QMC binding energy results of
H$_2$O overestimate the exact values by $0.4(1)$, $2.7(2)$, $4.4(6)$
mE$_{\rm h}$, for the STO-6G, cc-pVDZ, and DZ ANO basis,
respectively. The CCSD(T) method, which is highly accurate at this
geometry, has errors of $0.06$, $0.47$, $0.48$ mE$_{\rm h}$,
respectively.  In the TZ ANO basis, CCSD(T) and QMC yield binding
energies of 0.3675 and 0.373(1)\,E$_{\rm h}$, respectively.  

As mentioned, the DZ ANO and TZ ANO calculations of H$_2$O 
were for the geometry in
Ref.~\cite{dmrg_H2O_41}. Using CCSD(T), we verified that the
experimental H$_2$O geometry would lower the molecular total energy by
2.7~mE$_h$ and 4.0~mE$_h$ for the two basis sets. This would increase
the binding energy by the same amount, and bring the CCSD(T) binding
energy with the TZ ANO basis to good agreement with 0.370\,E$_{\rm
h}$, the basis extrapolated value of Feller and Peterson
\cite{Feller_Peterson}.  For comparison, the experimental binding
energy of H$_2$O is $0.3707$\,E$_{\rm h}$ (with the zero-point energy
removed) \cite{Feller_Peterson}.

Figure~\ref{fig:H2O_geom} shows a comparison of QMC and CCSD(T) 
results in the vicinity of the H$_2$O equilibrium bond length, using
cc-pVDZ and ANO double zeta basis sets. The angle between the two O-H
bonds is fixed at the experimental angle of 104.4798 degrees, and we
varied $R/R_e$ where $R_e=1.81$ Bohr is the experimental bond length
\cite{H2O_ref_exp}.  For the cc-pVDZ basis, the
CCSD(T) and QMC equilibrium bond lengths calculated from these curves
are 1.007~R$_e$ and 1.007(2)~R$_e$, respectively.  The 
corresponding values using the ANO DZ basis set are
1.003~R$_e$ and 1.006(4)~R$_e$.

\begin{table}[t]
\caption{The first ionization potential of first-row elements as
calculated in the cc-pVDZ and cc-pVTZ basis sets. For selected
elements, 
additional cc-pVQZ and cc-pV5Z basis sets are also used.  We show the
values obtained using HF, CCSD(T),  and QMC, together with
experimental results \cite{ion_book,andersen_ion}.  QMC statistical
errors are in the last digit and are shown in parentheses.  All
energies are in eV. }
\begin{tabular}{l c c c c}
\hline
\hline
Atom & HF & CCSD(T)&  QMC &  Exp \\
\hline
cc-pVDZ & \\
Li & $5.342 $  & $5.345 $  & $5.345(1) $  & $  5.39  $ \\
 Be & $8.079 $  & $9.290 $  &  $9.296(2) $  & $  9.32  $ \\
 B & $8.038 $  & $8.066 $  &  $7.942(5) $  & $  8.30  $ \\
 C & $10.803 $  & $10.983 $  &  $11.025(5) $  & $  11.26  $ \\
 O & $11.965 $  & $12.853 $  &  $12.808(2) $  & $  13.62  $ \\
 N & $13.895 $  & $14.195 $  &  $14.306(3) $  & $  14.53  $ \\
 F & $15.640 $  & $16.710 $  &  $16.725(8) $  & $  17.42  $ \\
 Ne & $19.668 $  & $20.893 $    & $20.948(5) $  & $  21.56  $ \\
\\
cc-pVTZ & \\
Li & $5.342 $  & $5.353 $    & $5.353(1) $  & $  5.39  $ \\
 Be & $8.045 $  & $9.285 $   & $9.259(7) $  & $  9.32  $ \\
 B & $8.038 $  & $8.228 $    & $8.08(1) $  & $  8.30  $ \\
 C & $10.798 $  & $11.184 $   & $11.237(8) $  & $  11.26  $ \\
 O & $12.011 $  & $13.326 $   & $13.256(7) $  & $  13.62  $ \\
 N & $13.892 $  & $14.449 $   & $14.574(4) $  & $  14.53  $ \\
 F & $15.654 $  & $17.142 $    & $17.154(7) $  & $  17.42  $ \\
 Ne & $19.673 $  & $21.309 $   & $21.37(1) $  & $  21.56  $ \\
\\
cc-pVQZ & \\
 B & $ 8.041 $  & $ 8.260 $    &  $ 8.19(5) $  & $  8.30  $ \\
O & $ 12.018 $  & $ 13.493 $    & $ 13.44(1) $  & $  13.62  $ \\
 F & $ 15.649 $  & $ 17.314 $    & $ 17.33(2) $  & $  17.42  $ \\
 Ne & $ 19.661 $  & $ 21.488 $   & $ 21.57(2) $  & $  21.56  $
 \\
\\
cc-pV5Z\\
 O & $ 12.022 $  & $ 13.558 $   &  $ 13.50(1) $ & $  13.62  $ \\
\hline 
\hline
\end{tabular}
\label{summary_ion1}
\end{table}

\begin{table}[t]
\caption{ Same as Table~\ref{summary_ion1}, but for 
the electron affinity and using the aug-cc-pVDZ   and
aug-cc-pVTZ basis sets \cite{aug_cc_basis}.}
\begin{tabular}{l c c c c}
\hline
\hline
Atom & HF & CCSD(T)& QMC &  Exp \\
\hline
aug-cc-pVDZ&\\
 B & $ -0.300 $  & $ 0.172 $  & $ 0.125(5) $  & $   0.28  $ \\
 C & $ 0.468 $  & $ 1.145 $   & $ 1.205(6) $  & $  1.263  $ \\
 O & $ -0.523 $  & $ 1.189 $   & $ 1.25(1) $  & $  1.46  $ \\
N & $ -1.858 $  & $ -0.512 $   & $ -0.52(1) $  & $  -0.07(2)  $ \\
 F & $ 1.284 $  & $ 3.228 $    & $ 3.40(1) $  & $  3.4  $ \\
 Ne & $ -7.724 $  & $ -7.321 $  & $ -7.32(1) $  &  \\
\\
aug-cc-pVTZ&\\
 B & $ -0.305 $  & $ 0.235 $ &$ 0.23(4) $  &   $   0.28  $ \\
 C & $ 0.453 $  & $ 1.226 $   &$ 1.29(3) $  &  $  1.263  $ \\
 O & $ -0.565 $  & $ 1.336 $ & $ 1.40(5) $  &  $  1.46  $ \\
 N & $ -1.813 $  & $ -0.296 $  &  $ -0.28(3) $  &$  -0.07(2)  $ \\
 F & $ 1.195 $  & $ 3.317 $  & $ 3.59(4) $  & $  3.4  $ \\
 Ne & $ -6.479 $  & $ -6.108 $   & $ -6.30(7) $  &  \\
\hline
\hline
\end{tabular}
\label{summary_ion2}
\end{table}

We present results for the O$_2$ molecule in
Table~\ref{table_O2_bind}.  The experimental equilibrium bond length
of $R_e=1.2074$\,\AA\, is used for all of these calculations.  Results
are shown for the three larger basis sets used in the above H$_2$O
calculations.  These include the cc-pVDZ and the DZ and TZ ANO basis
sets, which give 28, 46, and 92 basis functions for O$_2$,
respectively.  QMC and CCSD(T) results are again in good agreement.
In the TZ ANO basis, the CCSD(T) and QMC binding energies are 0.188
and 0.192(2)~E$_{\rm h}$, respectively. Our QMC value is in excellent
agreement with our previous binding energy 0.191(4) calculated using
pseudopotentials and planewaves \cite{alsaidi}.  Also for comparison,
the basis extrapolated binding energy at the CCSD(T) level is
0.191~E$_{\rm h}$ and the experimental value is 0.192~E$_{\rm h}$
(with the zero-point energy removed) \cite{Feller_Peterson}.

\subsection{First-row atoms and diatomic molecules}

\subsubsection{All-electron total ground-state energies}
\label{ssec:Other_1st_row_Etotal}

In Table~\ref{summary_molecules}, we show the total energies of
various first-row atoms and diatomic molecules at the equilibrium
geometry as calculated using RHF, UHF, CCSD(T), and also the FCI
energies where available.  For reference, we show also some available
E$_{\rm{est}}^\infty$, the estimated experimental non relativistic,
infinite nuclear mass energy
\cite{kolos_roothan,davidson_est_enyergy,filippi_umrigar}. These
represent the estimated exact results at the infinite basis size
limit, and are thus significantly lower than the CCSD(T), FCI, or QMC
energies because of the small basis size chosen in these benchmark
calculations.

For singlets  where both RHF and UHF
solutions exist, QMC calculations were done with each as a trial wave
function, and both results are shown in Table~\ref{summary_molecules}. In such
cases, QMC with the best variational single Slater determinant as
trial wave function, namely QMC/UHF, appears to always give better
energy values. We discuss this effect further in
Sec.\ref{sec:H2O-bondstretch} with stretched bonds in H$_2$O.

Our QMC energies are generally within a few mE$_h$  of
the FCI or the CCSD(T) energies. 
It  is interesting to note the cases of the berylium atom and molecule.
The Be atom, which has a near $2s$-$2p$ degeneracy, has often been used as
a benchmark in Green's function or diffusion Monte Carlo (DMC) 
studies \cite{Umrigar_DMC,Casula_geminal}. 
Optimized Slater-Jastrow trial wave functions with a single determinant are
known to lead to significant errors, 
of  $\sim 10$~mE$_h$, in the calculated fixed-node ground-state energy \cite{Umrigar_DMC}. 
(This error is largely removed when multi-determinant trial wave functions are used to
account for the near degeneracy. For example, DMC with an optimized Slater-Jastrow
trial wave function using
four determinants is extremely accurate \cite{Umrigar_DMC}.) 
In the present QMC method, the phaseless approximation using a single
Slater determinant appears to be significantly less dependent
on the trial wave function, 
with systematic errors of $\sim 1$~mE$_h$ or less.
Of course, the DMC calculations work in real configuration space and
thus have 
the advantage of no finite-basis
errors. Measured as a fraction of 
the correlation energy, the present QMC/UHF still has a
substantially smaller systematic error, of about 2\
Be/cc-pVTZ. 

Correspondingly, the Be$_2$ dimer is a notorious case in quantum chemistry \cite{Martin_CPL}.
A DMC calculation using optimized single-determinant trial wave functions did
not obtain binding \cite{Schautz_Be2-98}. Our QMC/UHF total energies are within $\sim 2$~mE$_h$ of CCSD(T), 
and the resulting binding energy with the larger cc-pVTZ basis 
is  $5.5(4)$~mE$_h$, in good agreement with the  CCSD(T) value of
$6.1$~mE$_h$.
The finite basis-size errors are still significant at this basis size ---
at the CCSD(T) level, the binding energy is $3.6$~mE$_h$ with the cc-pVQZ basis 
and $4.2$~mE$_h$ with the cc-pV5Z basis. Thus a more detailed study is
necessary before a direct and definitive comparison can be made between QMC
and experiment (BE: $4.0$~mE$_h$).
The  QMC result is consistent with  the previous  calculation using a plane-wave
basis \cite{zhang_krakauer}, which should be at the infinite basis limit
(aside from pseudopotential errors).

\subsubsection{Ionization potentials and electron affinities}
\label{sssec:IP_EA}

In Tables~\ref{summary_ion1} and \ref{summary_ion2} we present a
summary of the calculated first ionization potential and electron
affinity for first-row elements. For the ionization energy study, we
used the cc-pVDZ and cc-pVTZ double- and triple-zeta basis sets
\cite{cc_basis} for all the elements. A few selected elements are then
studied, using the larger cc-pVQZ basis sets. Also we looked at O with
a cc-pV5Z basis set with 91 basis functions. For the
electron affinity we used the aug-cc-pVDZ and aug-cc-pVTZ sets
\cite{aug_cc_basis}. For comparison, we show the values calculated  using
HF, CCSD(T), and QMC, as well as the experimental data  from
Refs.~\cite{ion_book,andersen_ion}. 

The agreement between QMC and the coupled cluster CCSD(T) results is
in general very good.  For unstable or meta-stable negative ions
(N$^-$ and Ne$^-$), the Hamiltonian expressed with a localized basis
always yields an atomic-like ground state. In
Tables~\ref{summary_ion1} and \ref{summary_ion2}, the total
ground-state energies are not shown, but the mean difference between
the QMC and CCSD(T) values among all the atoms and ions is less than
3~mE$_h$. The negative charged F ion, F$^-$/aug-cc-pVTZ, has the
largest discrepancy of $\sim 7$~mE$_h$. An MCSCF study \cite{nwchem}
showed that the RHF single determinant gives a rather poor description
of the ion.

\begin{table*}
\caption{Basis-size and trial wave function dependence in stretched-bond
calculations in H$_2$O. Equilibrium geometries are the same as in Table.~I.
The all-electron total energy of H$_2$O is shown using
  three different basis sets: minimal STO-6G, cc-pVDZ, and DZ ANO
  \cite{roos_footnote1} with 7, 24, and 41 bases functions,
  respectively. The O-H bond lengths are stretched to $1.5\,R_e$ and
  $2\,R_e$. QMC energies obtained with both RHF and UHF trial
  wave functions are shown in the last two columns.
 Statistical
  errors in QMC are in the last digit, and are shown in parentheses. 
FCI/DMRG results are from the same sources as in Table~\ref{table_H2O_bind}.
All energies are in Hartrees.
}
\begin{tabular}{l l l l l l l}
\hline
\hline
bond length  &  RHF &  UHF & CCSD(T)&  FCI  &  QMC/RHF  &  QMC/UHF \\
\hline
STO-6G   \\
$1.5\,R_e$ & -75.440\,432 &  -75.502\,069 &   &-75.600\,039 &-75.576\,8(3) &-75.596\,5(6) \\
$2\,R_e$  &  -75.141\,587 &  -75.464\,541 &   &-75.486\,528 &-75.355\,7(3) &-75.488\,0(3)  \\
\\
 cc-pVDZ\\
$1.5\,R_e$ &-75.802\,387 &  -75.829\,813 &-76.070\,717  &-76.072\,348 &  -76.068\,8(3) &-76.069\,7(2)  \\
$2\,R_e$  & -75.587\,711 &  -75.793\,668 &-75.955\,485  &-75.951\,665 &  -75.897\,3(4) &-75.954\,6(2) \\
\\
DZ ANO \\
$1.5\,R_e$ &  -75.817\,273 & -75.852\,670 & -76.129\,442  &
-76.131\,050
&  -76.126\,5(7) & -76.129\,0(5) \\
$2\,R_e$   &  -75.602\,850 &  -75.818\,969 &-76.009\,395 &     & -75.950(1) &-76.011(1) \\
\hline
\hline
\end{tabular}
\label{table_H2O_bond}
\end{table*}

\section{H$_2$O bond stretching}
\label{sec:H2O-bondstretch}
We next examine the accuracy of our method in describing bond stretching.
The ability of a computational method to deliver uniform accuracy as
bonds are stretched/broken is obviously important in chemistry. It
also provides an indicator for the potential accuracy of a method for
solids, mimicking different levels of electron correlation.  
Our method uses a trial wave function in the constraint to deal with the 
phase problem. Almost all calculations to date, both with the plane-wave basis and in
the present study, have used single Slater determinant
trial wave functions. Since the quality of such wave functions decreases 
as bonds are stretched and
correlation effects become more important, 
it is a significant challenge 
for the method to maintain its quality and obtain
uniformly accurate results. 
 
We apply our method to stretched bonds in H$_2$O, which has served as an
excellent benchmark system \cite{H2O_Olsen,dmrg_H2O_41}. 
Symmetric O-H bond length stretching is studied. The bond
angle between the two O-H bonds is held fixed, while the O-H bond
length is increased from its equilibrium value. We considered three different
basis sets: STO-6G, cc-pVDZ \cite{cc_basis}, and DZ ANO \cite{roos}. For the
small basis, we used the same geometry as in Table~I. For the two
larger basis sets, we used the same geometries as those of
Ref.~\cite{H2O_Olsen} and Ref.~\cite{dmrg_H2O_41} for
comparison with their FCI and DMRG energies, respectively. 

The results are shown in Table~\ref{table_H2O_bond}. 
In addition to QMC/UHF, i.e., using the variationally optimal HF solution as the trial wave function,
we also carried out QMC calculations using the RHF solution, in order to further examine 
the effect of the trial wave function.  As bonds are stretched,
static correlation becomes increasingly important.
The RHF solution becomes inceasingly unfavorable compared to the
UHF solution, which has the correct behavior in the dissociation limit.  
As can be seen in Table~\ref{table_H2O_bond}, QMC with the RHF trial
wave function (QMC/RHF) performs worse than QMC/UHF, which is consistent with the trend seen in
Table~\ref{summary_molecules}. Indeed, the QMC/RHF results deterioate as
the bonds are stretched, reflecting 
the qualitatively incorrect nature of the RHF trial wave function at large
bond lengths.

In the range of bond lengths in Table~\ref{table_H2O_bond},
QMC with the UHF trial wave function
gives roughly comparable accuracy as CCSD(T). 
For example, in the cc-pVDZ basis at $1.5\,R_e$, QMC/UHF
over-estimates the energy by 2.7(1)\,mE$_{\rm h}$, while
CCSD(T) over-estimates it by 1.6\,mE$_{\rm h}$.  At $2\,R_e$, both
QMC/UHF and CCSD(T) under-estimate the energy, by 3.0(2) and
3.8\,mE$_{\rm h}$, respectively.  In the DZ ANO basis for $1.5~R_e$, QMC/UHF is
above the DMRG value by $2.1(5)$~mE$_{\rm h}$, while
CCSD(T) is above by $1.6$~mE$_{\rm h}$.

\begin{figure}[t]
\includegraphics[width=8 cm]{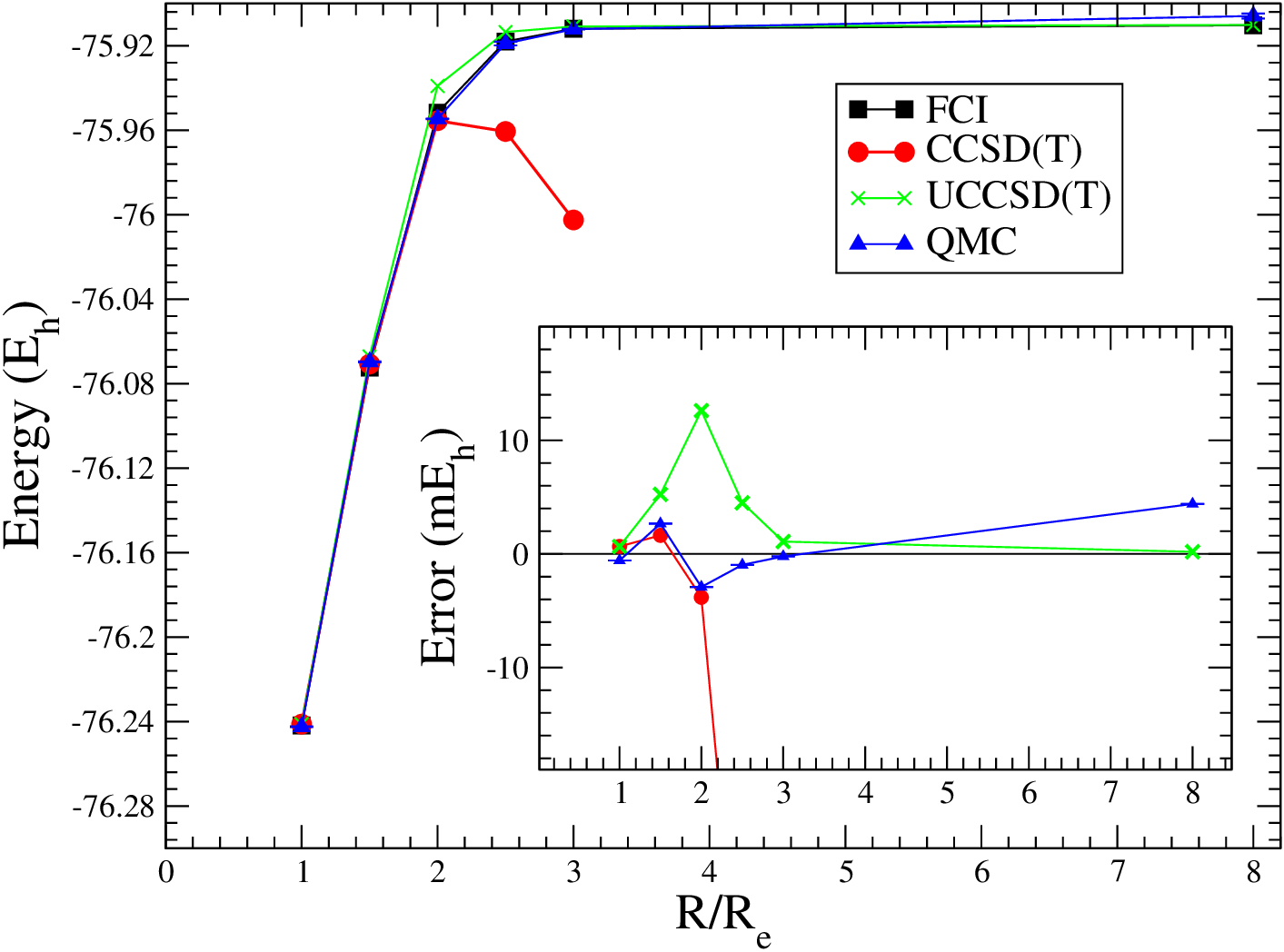} 
\caption{ 
A comparison of bond stretching in H$_2$O/cc-pVDZ between the present QMC, coupled
cluster methods, and FCI~\cite{H2O_Olsen}. The main graph shows the calculated
total energy (in Hartrees) as a function of the O-H bond length. The inset
shows the errors (in milli-Hartrees) of various methods with respect to FCI.
Points are connected by straight lines for clarity. QMC statistical error
bars are shown.}
\label{fig_H2O_stretch}
\end{figure}

Larger bond stretching of up to 8\,R$_e$ is presented in 
Figure~\ref{fig_H2O_stretch}, using the cc-pVDZ valence double-zeta basis. 
QMC results are
compared with exact FCI~\cite{H2O_Olsen} and with coupled cluster results 
using RHF and UHF reference states (CCSD(T) and UCCSD(T), respectively).
The inset shows 
the errors of the various methods from the FCI numbers. 
CCSD(T) is excellent near equilibrium, but is much worse at
large bond lengths. (As mentioned, QMC with  RHF trial wave functions would 
show similar behavior at large bond lengths.) UCCSD(T) performs much better for larger bond
lengths, but is worse in the intermediate regime with errors larger
than $10$\,mE$_{\rm h}$. 
Our QMC
results are in good agreement with the exact energies, showing a
maximum discrepancy of about 4(1)\,mE$_{\rm h}$, which occurs at $8R_e$.
Together with the equilibrium results, the method is seen to yield
rather uniform accuracy across these bond lengths. This is 
encouraging, given that it is achieved with the same choice,  
namely the variational UHF solution, as the trial wave function throughout the
entire region.

\section{Discussion and Summary}
\label{sec:summary}

The present QMC method provides an approximate, but non-perturbative approach 
for many-electron calculations. 
It can directly incorporate 
traditional electronic-structure machinery, 
such as high-quality basis sets, effective-core potentials, etc.  
The method obtains the many-body ground state
by building 
a stochastic linear superposition of non-orthogonal 
independent-particle solutions. 
Thus, algorithmically it shares many of the ingredients in more standard
methods in quantum chemistry and solid state physics.
Its computational cost scales with the number
of basis functions as $N^3$ to  $N^4$. 

The
accuracy of the present QMC method depends on the reference or trial
wave function $\left| \Psi_T \right\rangle$, which in this study is chosen as the HF
state.  As in coupled cluster methods, the trial wave function is the starting
point of the ground-state projection, but in QMC all excitations are
implicitly included by the projection.  Errors arise from the
phaseless approximation in the projection, which uses $|\Psi_T\ra$ to
approximate the phase of the contribution of a Slater determinant to
the ground state, via the importance-sampling transformation.

We have found, as shown in Tables \ref{summary_molecules} and
\ref{table_H2O_bond}, that the most accurate energies are in general
obtained using the best variational single determinant 
$|\Psi_T\ra$. This is simply the HF solution when RHF and UHF are the
same (e.g., in the H$_2$O molecule at equilibrium), and is the UHF
solution when the two differ. In the latter case, the method seems
relatively insensitive, within the unrestricted framework, to whether
a Hartree-Fock or hybrid B3LYP Slater determinant is used.

In contrast to the diffusion Monte Carlo (DMC) method, the present QMC must
deal with finite basis-set errors and basis-size convergence. 
However, the ability to use any
single-particle basis can also be advantageous. 
For example, in addition to the connection to quantum chemistry methods,  
our new method also makes possible QMC calculations for general
model Hamiltonians, which are frequently
used in the study of correlated electron systems. 
Like the present AF QMC method, DMC is also approximate, using a trial wave
function to impose the 
fixed-node approximation to control the sign problem. 
The trial wave functions that have been used in our method are much simpler.
Because of the complementary nature of the two different methods, direct 
comparisons are not always straightforward.
Indications from the current calculations 
and earlier plane-wave studies are that the systematic accuracy
of the phaseless approximation compare favorably with
fixed-node DMC. 

In summary, we extended the recently introduced phaseless QMC method
to handle any one-particle basis, and applied it to atoms and
molecules using Gaussian basis sets. Overall, our results at and near
the equilibrium geometries are roughly comparable to those obtained
using CCSD(T), but are superior for bond stretching.  Our preliminary
results (to be published elsewhere) on bond-breaking in several
diatomic molecules show similar uniform accuracy.  For a first
application, these results are quite encouraging. There are many
possibilities for further improvement of the method.  Currently, we
are investigating different forms of Hubbard-Stratonovich
transformations, possibilities for better scaling, as well as
applications of the present method in transition metal oxides and
other systems.

\section{Acknowledgments:}

We would like to thank E.~J.~Walter and S.~R.~White for useful discussions. This
work is supported by ONR (N000149710049, N000140110365, and N000140510055),
NSF (DMR-0535529), and ARO (48752PH) grants, and by the DOE 
computational materials science network (CMSN).  
Computations were carried out at
the Center for Piezoelectrics by Design, the SciClone Cluster at the
College of William and Mary, NCSA at UIUC, and SDSC at UCSD.


\begin{thebibliography}{10}
\bibitem{qc_ref} A.~Szabo and N.~S.~Ostlund, \emph{Modern Quantum
Chemistry} (McGraw-Hill, New York, 1989).

\bibitem{white_1}  S. R. White and R. L. Martin, J. Chem. Phys. {\bf 110},
  4127 (1999).

\bibitem{dmrg_H2O_41} Garnet Kin-Lic Chan and Martin Head-Gordon,
J.~Chem. Phys. {\bf 118}, 8551 (2003).

\bibitem{dmrg_rev} U.~Schollw\"ock, Rev. Mod. Phys. {\bf 77}, 259 (2005).

\bibitem{ccsdt_ref} J. Cizek, J. Chem. Phys. {\bf 45}, 4256 (1966).

\bibitem{ccsdt_crawford} T. Crawford and H. Schaefer, Rev. Comp. Chem. {\bf 14}, 33 (2000). 


\bibitem{QMC_rmp} W.~M.~C.~Foulkes, L.~Mitas, R.~J.~Needs, and
G.~Rajagopal, Rev. Mod. Phys. {\bf 71}, 33 (2001).

\bibitem{QMC_rev_mole} D.~M.~Ceperley and L.~Mitas, 
in {\it New Methods in Computational Quantum
  Mechanics Advances in Chemical Physics, XCIII\/}, eds.~I.~Prigogine and
  S.~A.~Rice, 1996;  
  B.~L.~Hammond, W.~A.~Lester, Jr. and P.~J.~Reynolds,
 {\it Monte Carlo methods in ab initio quantum chemistry\/}
  (World Scientific, Singapore, 1994).

\bibitem{zhang_krakauer} Shiwei Zhang and Henry Krakauer, Phys. Rev. Lett. {\bf 90},
  136401 (2003).

\bibitem{BSS} R.~Blankenbecler, D. J. Scalapino, and R. L. Sugar,
Phys.\ Rev.\ D {\bf 24}, 2278 (1981).


\bibitem{Koonin} G.~Sugiyama and S.~E.~Koonin,
Ann.\ Phys.\ (NY) {\bf 168}, 1 (1986).

\bibitem{Zhang_review} Shiwei Zhang, in 
{\it Theoretical Methods for Strongly Correlated Electrons\/},
   ed.~by D. Senechal, A.-M. Tremblay, and C. Bourbonnais (Springer 2003).


\bibitem{Zhang} Shiwei Zhang, J.~Carlson, and J.~E.~Gubernatis, Phys.\
Rev.\ B {\bf 55}, 7464 (1997).

\bibitem{Silvestrelli93} P. L. Silvestrelli, S. Baroni, and R. Car, 
Phys.\ Rev.\ Lett.\ {\bf 71}, 1148 (1993).

\bibitem{shiftcont_rom97} Naomi Rom, D.~M.~Charutz, and Daniel
Neuhauser, Chem.\ Phys.\ Lett. \textbf{270}, 382 (1997).

\bibitem{Wilson95}M.~T.~Wilson and B.~L.~Gyorffy,
  J.~Phys.~Condens.~Matter \textbf{7}, 371 (1995).


\bibitem{zhang_krakauer2} Shiwei Zhang, Henry Krakauer, Wissam  Al-Saidi,
  and Malliga Suewattana, Comp. Phys. Comm. {\bf 169}, 394 (2005).

\bibitem{cherry} Malliga Suewattana \etal  (in preparation). 

\bibitem{alsaidi} W. A. Al-Saidi, Henry Krakauer, and Shiwei Zhang, Phys. Rev. B
{\bf 73}, 075103 (2006).
  
\bibitem{roos} {P. O. Widmark, P. A. Malmqvist, and B. Roos,
Theo. Chim. Acta, 77, 291 (1990).}

\bibitem{roos_footnote1} Basis sets can be obtained from the
Extensible Computational Chemistry Environment Basis Set Database
(http://www.emsl.pnl.gov/forms/basisform.html). The ANO basis sets
used for H$_2$O and O$_2$ are designated ``Roos Augmented Double Zeta ANO'' and
``Roos Augmented Triple Zeta ANO.''

\bibitem{cc_basis} { Thom H. Dunning, J. Chem. Phys. {\bf 90}, 1007 (1989).}

\bibitem{HS} R.~L.~Stratonovich, Sov.\ Phys.\ Dokl. \textbf{2},
416 (1958); J.~Hubbard, Phys.~Rev.~Lett.
{\bf 3}, 77 (1959).

\bibitem{Wirawan05}Wirawan Purwanto and Shiwei Zhang,
Phys.~Rev.~A {\bf 72}, 053610 (2005).

\bibitem{Wirawan04}Wirawan Purwanto and Shiwei Zhang,
Phys.~Rev.~E {\bf 70}, 056702 (2004).

\bibitem{gamess} M.~W.~Schmidt, K.~K.~Baldridge, J.~A.~Boatz, S.~T.~Ebert, M.~S.~Gordon,
J.~H.~Jensen, S.~Koseki, N.~Matsunaga, K.~A.~Nguyen, S.~J.~Su, T.~L.
Windus, M.~Dupuis, and J.~A.~Montgomery, J.~Comput.~Chem.~{\bf 14},
1347 (1993); http://www.msg.ameslab.gov/GAMESS/GAMESS.html.


\bibitem{nwchem}{
 T.~P.~Straatsma, E.~Apr\'a, T.~L.~Windus, E.~J.~Bylaska, W.~de Jong,
            S.~Hirata, M.~Valiev, M.~T.~Hackler, L.~Pollack, R.~J.~Harrison,
            M.~Dupuis, D.~M.~A.~Smith, J.~Nieplocha, V.~Tipparaju,
            M.~Krishnan, A.~A.~Auer, E.~Brown, G.~Cisneros, G.~I.~Fann,
            H.~Fruchtl, J.~Garza, K.~Hirao, R.~Kendall, J.~A.~Nichols,
            K.~Tsemekhman, K.~Wolinski, J.~Anchell, D.~Bernholdt, P.~Borowski,
            T.~Clark, D.~Clerc, H.~Dachsel, M.~Deegan, K.~Dyall, D.~Elwood,
            E.~Glendening, M.~Gutowski, A.~Hess, J.~Jaffe, B.~Johnson, J.~Ju,
            R.~Kobayashi, R.~Kutteh, Z.~Lin, R.~Littlefield, X.~Long, B.~Meng,
            T.~Nakajima, S.~Niu, M.~Rosing, G.~Sandrone, M.~Stave, H.~Taylor,
            G.~Thomas, J.~van Lenthe, A.~Wong, and Z.~Zhang,
            "NWChem, A Computational Chemistry Package for Parallel Computers, 
            Version 4.6" (2004),
                      Pacific Northwest National Laboratory,
                      Richland, Washington 99352-0999, USA.}


\bibitem{molpro} MOLPRO version 2002.6 is a package of ab initio programs written by
H.-J.~Werner and P.~ J.~Knowles, with contributions from J.~Almlof,
R.~D.~Amos, M.~J.~O.~ Deegan, S.~T.~Elbert, C.~Hampel, W.~Meyer,
K.~A.~Peterson, R.~M.~Pitzer, ¨ A.~J.~Stone, P.~R.~Taylor, and
R.~Lindh, Universitat Bielefeld, Bielefeld, Germany, University of
Sussex, Falmer, Brighton, England, 1996.


\bibitem{fci_molpro} P.~J.~Knowles and N.~C.~Handy,
Chem.~Phys.~Letters 111, 315 (1984); P.~J.~Knowles and N.~C.~Handy,
Comp.~Phys.~Commun.~54, 75 (1989).

\bibitem{saxe} P.~Saxe, H.~F.~Schaefer III, and N.~C.~Handy,
 Chem. Phys. Lett. {\bf 79}, 202 (1981).

\bibitem{Bauschlicher} C.~W.~Bauschlicher and P.~R.~Taylor,
J. Chem. Phys. {\bf 85}, 2779 (1986).

\bibitem{H2O_Olsen} J.~Olsen, P.~Jorgensen, H.~Koch, A.~Balkova, and
R.~J.~Bartlett, J.~Chem. Phys. {\bf 104}, 8007 (1995).

\bibitem{H2O_DMC} A.~Luchow, J.~B.~Anderson, and D.~Feller,
J. Chem. Phys. {\bf 106}, 7706 (1997).

\bibitem{Feller_Peterson} D.~Feller and K.~A.~Peterson,
J. Chem. Phys. {\bf 110}, 8384 (1999).

\bibitem{H2O_ref_exp} A.~R.~Hoy and P.~R.~Bunker, J. Molecular Structure {\bf
74}, 1, (1979).

\bibitem{kolos_roothan} W. Kolos and C. C. J. Roothaan, Rev. Mod. Phys. {\bf 32},  219 (1960).

\bibitem{davidson_est_enyergy} Ernest R. Davidson, Stanley A. Hagstrom,
  Subhas J. Chakravorty, Verena Meiser Umar, and Charlotte Froese
  Fischer, Phys. Rev. A {\bf 44}, 7071 (1991). 

\bibitem{filippi_umrigar} Claudia Filippi and C. J. Umrigar, J. Chem. Phys. {\bf 105}, 213 (1996).


\bibitem{abrams} Micah L. Abrams and C. David Sherrill,
J. Chem. Phys. {\bf 118}, 1604 (2003).



\bibitem{Umrigar_DMC}C.~J.~Umrigar, M.~P.~Nightingale, and K.~J.~Runge,
J.\ Chem.\ Phys.\ {\bf 99}, 2865 (1993). 

\bibitem{Casula_geminal}M.~Casula and S.~Sorella, J.\ Chem.\
  Phys. {\bf 119}, 6500 (2003). 

\bibitem{Martin_CPL}Jan M.~L.~Martin,  Chem. Phys. Lett. {\bf303},
  399(1999). 

\bibitem{Schautz_Be2-98} F. Schautz, H.-J. Flad, and M. Dolg, Theor. Chem. Acc. 
{\bf 99}, 231 (1998).
  
\bibitem{ion_book} A.~A.~Radzig and  B.~M.~Smirnov, \emph{Reference data
  on atoms, molecules, and ions} (Springer-Verlag, 1985).

\bibitem{andersen_ion} T.~Andersen, H.~K.~Haugen, and H.~Hotop,
  J. Phys. Chem. Ref. Data {\bf 28}, 1511 (1999). 

\bibitem{aug_cc_basis} {Rick A. Kendall, Thom H. Dunning, Jr., and  Robert
  J. Harrison, J. Chem. Phys. {\bf 96}, 6796 (1992).}

\end{thebibliography}
\end{document}